\documentclass[twocolumn,runningheads]{svjour2}
\smartqed  % flush right qed marks, e.g. at end of proof
\usepackage{graphicx}
\journalname{Astrophysics and Space Science}
\begin{document}

\title{Gamma rays from molecular clouds%\thanks{Grants or other notes
}


\author{Stefano Gabici         \and
        Felix A. Aharonian     \and
	Pasquale Blasi
}


\institute{S. Gabici, F. A. Aharonian \at
              Max-Planck-Institut f\"ur Kernphysik, Saupfercheckweg 1, 69117 Heidelberg, Germany \\
              \email{Stefano.Gabici@mpi-hd.mpg.de}           %  \\
           \and
           P. Blasi \at
              INAF/Osservatorio Astrofisico di Arcetri, Largo Fermi 5, 50125 Firenze, Italy
}

\date{Received: date / Accepted: date}

\maketitle

\begin{abstract}
It is believed that the observed diffuse gamma ray emission from the galactic plane is the result of interactions between cosmic rays and the interstellar gas. Such emission can be amplified if cosmic rays penetrate into dense molecular clouds. The propagation of cosmic rays inside a molecular cloud has been studied assuming an arbitrary energy and space dependent diffusion coefficient. If the diffusion coefficient inside the cloud is significantly smaller compared to the average one derived for the galactic disk, the observed gamma ray spectrum appears harder than the cosmic ray spectrum, mainly due to the slower penetration of the low energy particles towards the core of the cloud. This may produce a great variety of gamma ray spectra.
\PACS{First \and Second \and More}
\end{abstract}

\section{Introduction}

The observed diffuse gamma ray emission from the galactic plane is believed to be the result of the decay of neutral pions produced during inelastic collisions of cosmic rays with interstellar gas. If, as a first order approximation, one assumes that the cosmic ray spectrum is the same everywhere in the Galaxy, then the gamma ray emission is expected to be simply proportional to the gas column density. As a consequence, if the line of sight is intersecting regions of enhanced density, such as giant molecular clouds, also the gamma ray emission is expected to be correspondingly enhanced.

The importance of the detection of molecular clouds in gamma rays is widely recognize, especially in relation to the problem of the origin of cosmic rays. Molecular clouds located in the vicinity of cosmic ray accelerators could provide a dense target for cosmic rays interactions, amplifying the resulting gamma ray emission and making easier the identification of cosmic ray sources \cite{montmerle,casse,felix1,felix2}.

On the other hand, even in the absence of an accelerator, molecular clouds embedded in the ``sea'' of galactic cosmic rays are expected to emit gamma rays. If cosmic rays can freely penetrate the clouds, the high energy gamma ray spectrum is expected to mimic the slope of the cosmic ray spectrum and the total gamma ray luminosity depends only on the cloud total mass. For this reason, molecular clouds can be used to probe the cosmic ray energy spectrum and its absolute flux in different parts of the Galaxy \cite{issa,felix1,felix2}. The condition for the detectability of these \textit{passive} clouds with EGRET is $M_5/d_{kpc} \ge 10$, where $M_5$ is the cloud mass in units of $10^5 M_{\odot}$ and $d_{kpc}$ is the source distance in kpc \cite{felix2}. Since there are only a few clouds in the galaxy with such a large value, a more sensitive instrument like GLAST is needed for these studies.

Of course, what discussed above is valid only under the assumption that cosmic rays freely penetrate the clouds. The issue of the penetration or exclusion of cosmic rays from clouds has been investigated in several papers \cite{heinz,strong,dogiel}, in which quite different conclusions have been drawn, going from the almost-free penetration to the exclusion of cosmic rays up to tens of GeV. Since a theoretical determination of the cosmic ray diffusion coefficient is a very difficult task, we adopt here a more phenomenological approach: the diffusion coefficient is parametrized, and observable quantities (such as the gamma ray flux) capable of constraining it are proposed.

In this paper we consider only \textit{passive} clouds embedded in the diffuse galactic cosmic ray flux and located far away from cosmic ray accelerator. The case of clouds interacting with accelerators will be considered in a forthcoming paper.

\section{Charcateristic time scales of the problem}

Before solving the equation that describes the transport of cosmic rays inside a magnetized, dense cloud, it is worth giving an estimate of the typical time scales involved in the problem. Consider a giant molecular cloud of radius $R_{cl} \sim 20 ~ {\rm pc}$, mass $M_{cl} \sim 2 ~ 10^5 M_{\odot}$ and average magnetic field $B_{cl} \sim 10 ~ \mu {\rm G}$. The average density (Hydrogen atoms) of such cloud is thus $n_{gas} \sim 300 ~ {\rm cm^{-3}}$, which gives a dynamical (free--fall) time for the system of the order of:
\begin{equation}
\tau_{dyn} \sim (G \rho )^{-\frac{1}{2}} \sim 5.5 \, 10^6 \left( \frac{n_{gas}}{300 \, {\rm cm^{-3}}}\right)^{-\frac{1}{2}} ~ {\rm yr}
\end{equation}
where $G$ is the gravitational constant and $\rho$ the mass density.
This free--fall time has probably to be considered as a strict lower limit to the lifetime of the cloud, since additional pressure support from fluid turbulence and magnetic field may inhibit the collapse \cite{review}. 

The effectiveness of the cosmic ray penetration into the cloud depends on the interplay of several physical processes: (i) diffusion in the cloud magnetic field, (ii) advection due to turbulent bulk motion inside the cloud, (iii) energy losses in the dense cloud medium.
Moreover, the cosmic ray density can be enhanced if a cosmic ray accelerator is embedded in the cloud \cite{felix1}, or if cosmic rays coming from outside the cloud are reaccelerated via Fermi--like processes that may take place in the magnetized cloud turbulence \cite{dogiel,dogiel2}. 
In the following we consider the cloud as a passive target for galactic cosmic rays and neglect any effect related to the possible presence of acceleration and/or reacceleration of cosmic rays inside the cloud.

We parametrized the diffusion coefficient for protons of energy $E$ propagating in the cloud magnetic field $B$ in the following way:
\begin{equation}
\label{diff}
D(E) = \chi D_0 \left( \frac{E/{\rm GeV}}{B/3 \mu {\rm G}} \right)^{\delta} 
\end{equation}
where $D_0 = 3 ~ 10^{27} {\rm cm^2/s}$ and $\delta = 0.5$ are the typical galactic values \cite{CRbook} and the parameter $\chi < 1$ accounts for a possible suppression of the diffusion coefficient inside the turbulent cloud medium. In general, the values of $\chi$ and $\delta$ will depend on the power spectrum of the magnetic field turbulence.
For such a choice of parameters, one can estimate the proton diffusion time, namely, the time it takes a proton to penetrate into the core of the cloud:
\begin{equation}
\tau_{diff} = \frac{R_{cl}^2}{6 D(E)} 
\end{equation}
$$
\sim 1.2 ~ 10^4 \chi^{-1} \left(\frac{R_{tot}}{20 ~ pc}\right)^2 \left( \frac{E}{GeV} \right)^{-0.5} \left( \frac{B}{10 \mu G} \right)^{0.5} ~ yr
$$
To study the effective propagation of cosmic rays into cluods, it is instructive to compare the diffusion time with the energy loss time. In the dense cloud environment, cosmic ray protons suffer energy losses due to ionization and nuclear p--p interactions. Above the energy threshold for pion production $E_{th} \approx 300 {\rm MeV}$ nuclear interactions dominate. Since both the cross section $\sigma_{pp} \sim 40$ mb and inelasticity $\kappa \sim 0.45$ of this process are not changing significantly over a broad range of proton energies from $\sim 1 ~ {\rm GeV}$ to hundreds of TeVs, the proton lifetime is almost energy independent:
\begin{equation}
\label{losst}
\tau_{pp} = \frac{1}{n_{gas} c \kappa \sigma_{pp}} \sim 2 ~ 10^5 \left(\frac{n_{gas}}{300 ~ {\rm cm^{-3}}}\right)^{-1} ~ {\rm yr}
\end{equation}

Cosmic rays can also be transported by the fluid turbulence which is known to be present in molecular clouds. Several molecular lines are observed in the direction of clouds, and their width $\Delta v$ reflects the velocity of internal turbulent motions. The line width is known to correlate with the cloud size according to the relation: $\Delta v \propto R_{cl}^{0.5} $ \cite{review}, which for the cloud sizes considered here provides a velocity of a few km/s. The time scale for this advective transport can be roughly estimated as:
\begin{equation}
\tau_{adv} \sim \frac{R_{cl}}{\Delta v} \sim 4 ~ 10^6 \left(\frac{R_{cl}}{20 ~ {\rm pc}}\right) \left(\frac{\Delta v}{5 {\rm km/s}}\right) ~ {\rm yr}
\end{equation}

\begin{figure*}
\centering
  \includegraphics[width=0.45\textwidth]{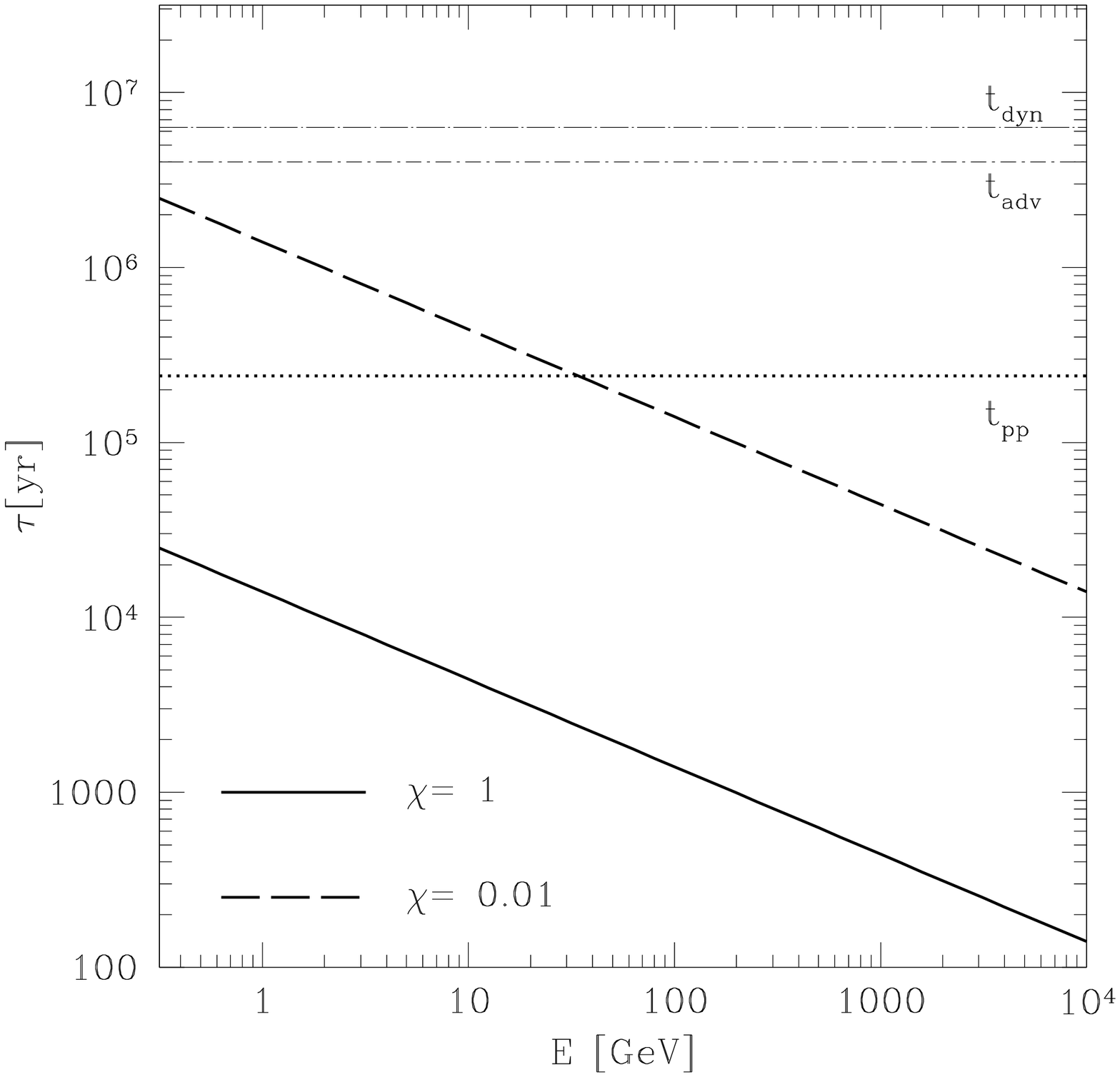}%{timeprotons.ps}
  \hspace{0.5cm} 
  \includegraphics[width=0.45\textwidth]{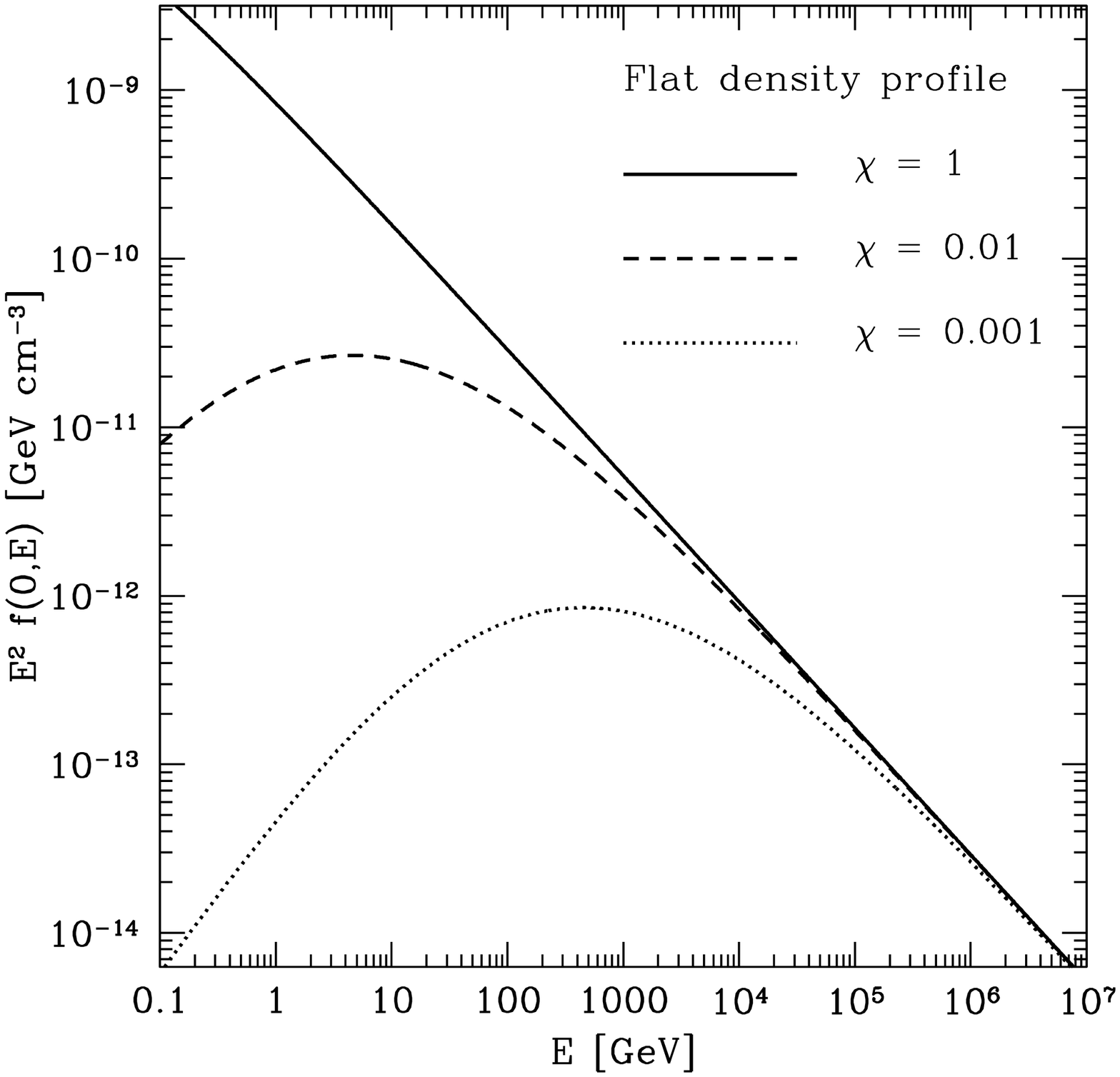}%{pspec0BW.ps}
\caption{{\bf LEFT PANEL}: typical time scales for cosmic rays in a giant molecular cloud with average number density $\sim 300 ~ \rm{cm}^{-3}$, radius 20 pc and average magnetic field 10 $\mu$G. Solid and dashed lines represent the diffusion time, horizontal lines represent the dinamical, advective and energy loss times (top to bottom). {\bf RIGHT PANEL}: cosmic ray spectrum in the cloud center for different values of the parameter $\chi$, describing the suppression of the diffusion coefficient with respect to the measured galactic value (see text for details).}
\label{fig:1}       % Give a unique label
\end{figure*}

The energy dependence of all the time scales considered above is shown in Figure \ref{fig:1} (left panel), where the solid and dashed thick lines represent the diffusion time (with $\chi = 1$ and $\chi = 0.01$ respectively), while the horizontal lines refers to the dynamical, advective and energy loss time scales (from top to bottom).
Several comments are in order. First of all, the dynamical lifetime of the cloud is the longest time scale for all the relevant energies. This means that it is possible to search for a steady state solution of the problem. Second, the advection time is comparable with the dynamical time, but it is always significantly longer than both the diffusion and the energy loss time scales. Thus, we can safely ignore the advection term in the cosmic rays transport equation.
As a consequence, the degree of penetration of cosmic rays inside the cloud can be roughly estimated by comparing the diffusion and energy loss times.
To this purpose, in Figure \ref{fig:1} (left panel) two different values of the parameter $\chi$ are considered. If $\chi = 1$ the diffusion time is shorter than the energy loss time at all the considered energies. Thus, if the diffusion coefficient inside the cloud is not suppressed with respect to the galactic value, cosmic rays can easily penetrate into the cloud. On the other hand, if diffusion inside the cloud is significantly suppressed ($\chi = 0.01$), the energy loss time becomes shorter than the diffusion time at energies below $E_{*} \sim 10 - 100 {\rm GeV}$. 
This means that only cosmic rays with energy above $E_{*}$ can penetrate into the cloud before losing their energy. This is a very improtant fact, since as we will demonstrate in the following sections, the exclusion of low energy cosmic rays plays a crucial role in shaping the gamma ray spectrum of molecular clouds.

Finally, it is worth stressing that the value of $E_{*}$ increases if one considers a realistic density profile for the cloud instead of average quantities. Despite the fact that density profiles cannot be easily extracted from available observations, it is well known that molecular clouds contain cores of size $\sim 1 ~ {\rm pc}$ or less, in which density can reach very high values $n_{gas} \sim 10^5 {\rm cm^{-3}}$. In this case, a proton which is approaching the center of the cloud meets a denser and denser environment. The value of the magnetic field is also increasing towards the cloud center, reaching in the densest regions values of hundreds of $\mu G$ or even more \cite{zeeman}. These facts make the energy loss time shorter and the diffusion time longer, leading to a more efficient exclusion of cosmic rays from clouds cores.

\section{Solution of the transport equation}

As shown above, the equation describing the cosmic ray transport in a cloud is the steady state diffusion--losses equation, which in spherical simmetry reads:
\begin{equation}
\label{diffloss}
\frac{1}{R^2}\frac{\partial}{\partial R} \left[ D(R,E) R^2 \frac{\partial f}{\partial R} \right] + \frac{\partial}{\partial E} \left[ \dot{E}(R) f \right] = 0
\end{equation}
where $f(R,E)$ is the space and energy dependent particle distribution function, $D(R,E)$ is the diffusion coefficient parametrized as described in the previous section and $\dot{E}(R,E) = dE/dt$ represents energy losses.

We parametrize the density profile as follows:
\begin{equation}
n_H(R) = \frac{n_0}{1+\left(\frac{R}{R_c}\right)^\alpha}
\end{equation}
where $n_0$ is the central density and $R_c$ the core radius, assumed to be 1/2 parsec. This is of course a simplified assumption, since molecular clouds can exhibit very irregular density profiles. However, we will show in the following that the mechanism of cosmic ray exclusion from clouds can work for very different density profiles, from flat ($\alpha = 0$) to very peaked ones ($\alpha \gg 0$). %This suggests that the same effect may be present also in more realistic (and complicated) situations.

In calculating the energy loss term we consider only inelastic proton--proton collisions, since this is the dominant process above the threshold for the production of pions ($E_{th} \sim 280$ MeV). The loss term in Eq. (\ref{diffloss}) depends on the density profile through Eq. (\ref{losst}), and thus it is space dependent.

For the magnetic field profile we use the results from \cite{zeeman}, in which Zeeman measurements of magnetic field strength in molecular cloud cores are reported. A correlation between magnetic field strength and gas density is observed and can roughly be fitted by:
\begin{equation}
\label{Bfield}
B \sim 100 \left( \frac{n_H}{10^4 cm^{-3}} \right)^{1/2} \mu G
\end{equation}
Though this correlation has been found for molecular cloud cores with density exceeding $n_H \sim 10^3 {\rm cm^{-3}}$, it provides reasonable values also for low density regions (tens of $\mu G$ for typical average cloud densities of a few $100 cm^{-3}$). Thus, we assume that Eq. (\ref{Bfield}) is valid in the entire density interval. It is worth noticing that in our model the diffusion coefficient is space--dependent, since it depends on the magnetic field as given by Eq. (\ref{diff}).

We solved Eq. (\ref{diffloss}) numerically, using an implicit scheme and assuming, as boundary condition, that the cosmic ray spectrum outside the cloud must match the galactic cosmic ray spectrum \cite{dogiel}. The galactic spectrum is in turn assumed to be equal to the locally observed cosmic ray flux:
\begin{equation}
J_{CR}^{gal}(E) = 2.2 \left(\frac{E}{GeV}\right)^{-2.75} cm^{-2} s^{-1} sr^{-1} GeV^{-1} 
\end{equation}

The effective exclusion of cosmic rays from the cloud cores is demonstrated in Fig. \ref{fig:1} (right panel), where the cosmic ray spectrum in the cloud center is plotted for different values of the parameter $\chi$ (here $\delta$ is set equal to the galactic value 0.5). In obtaining the result, a cloud with mass $2 \times 10^5 M_{\odot}$, radius 20 pc and a flat density profile (this correspond to a spatially constant magnetic field of $\sim 15 \mu$G) has been considered. If the diffusion coefficient inside the cloud is not suppressed with respect to the galactic value ($\chi = 1$), then the cosmic ray spectrum in the cloud center is basically indistinguishable from the galactic cosmic ray spectrum. On the other hand, is diffusion is significantly suppressed ($\chi << 1$), cosmic rays with energy above $\sim 10$ GeV ($\chi = 0.01$) or $\sim 100$ GeV ($\chi = 0.001$) cannot penetrate the cloud. As we will show in the following, this fact has important implications for the estimate of the spectrum and intensity of the gamma ray emission expected from molecular clouds.

Cosmic ray protons propagating inside a molecular cloud also produce secondary electrons during inelastic interactions in the intercloud medium. These electrons contribute to the overall gamma ray emission of the cloud via Bremsstrahlung. Once the steady state proton spectrum has been obtained, we calculated the injection spectrum $Q_e(E_e)$ of the secondary electrons by using the analytical fits provided in \cite{kelner}. The steady state spectrum of secondary electrons can be obtained using again Eq. (\ref{diffloss}) appropriately modified as follows: \textit{i)} the injection term $Q_e(E_e)$ must be added on the left side; \textit{ii)} the loss term $\dot{E}_e(R,E)$ is now dominated by Coulomb, Bremsstrahlung and synchrotron losses \cite{ginzburg} and it is both space and energy dependent; \textit{iii)} as boundary condition we assumed that the particle distribution function for secondary electron vanishes outside the cloud, where the gas density is low with inefficient production of secondaries. 

\section{Gamma ray spectra}

\begin{figure*}
\centering
  \includegraphics[width=0.45\textwidth]{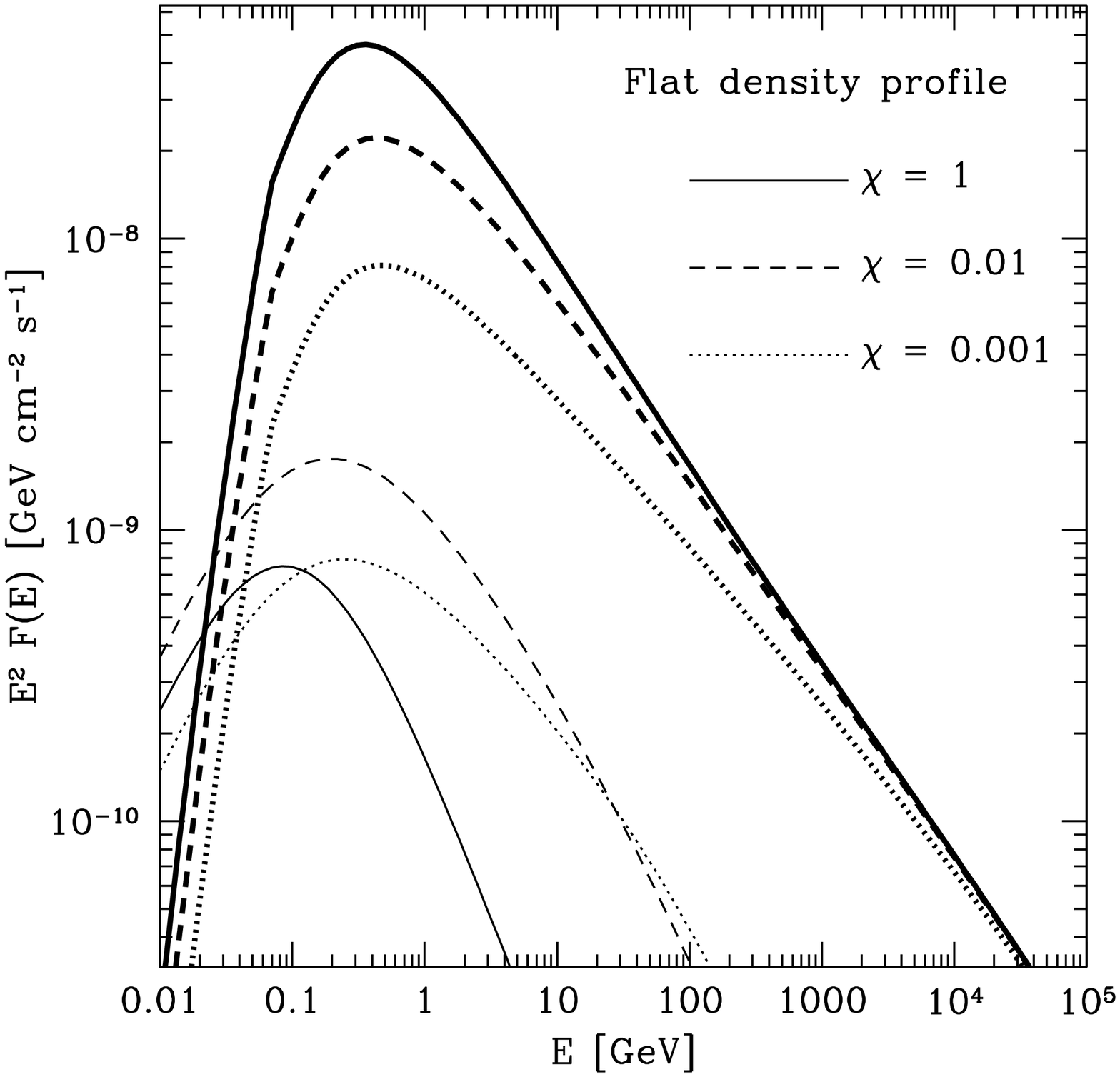}%{gammaBW.ps}
  \hspace{0.5cm} 
  \includegraphics[width=0.45\textwidth]{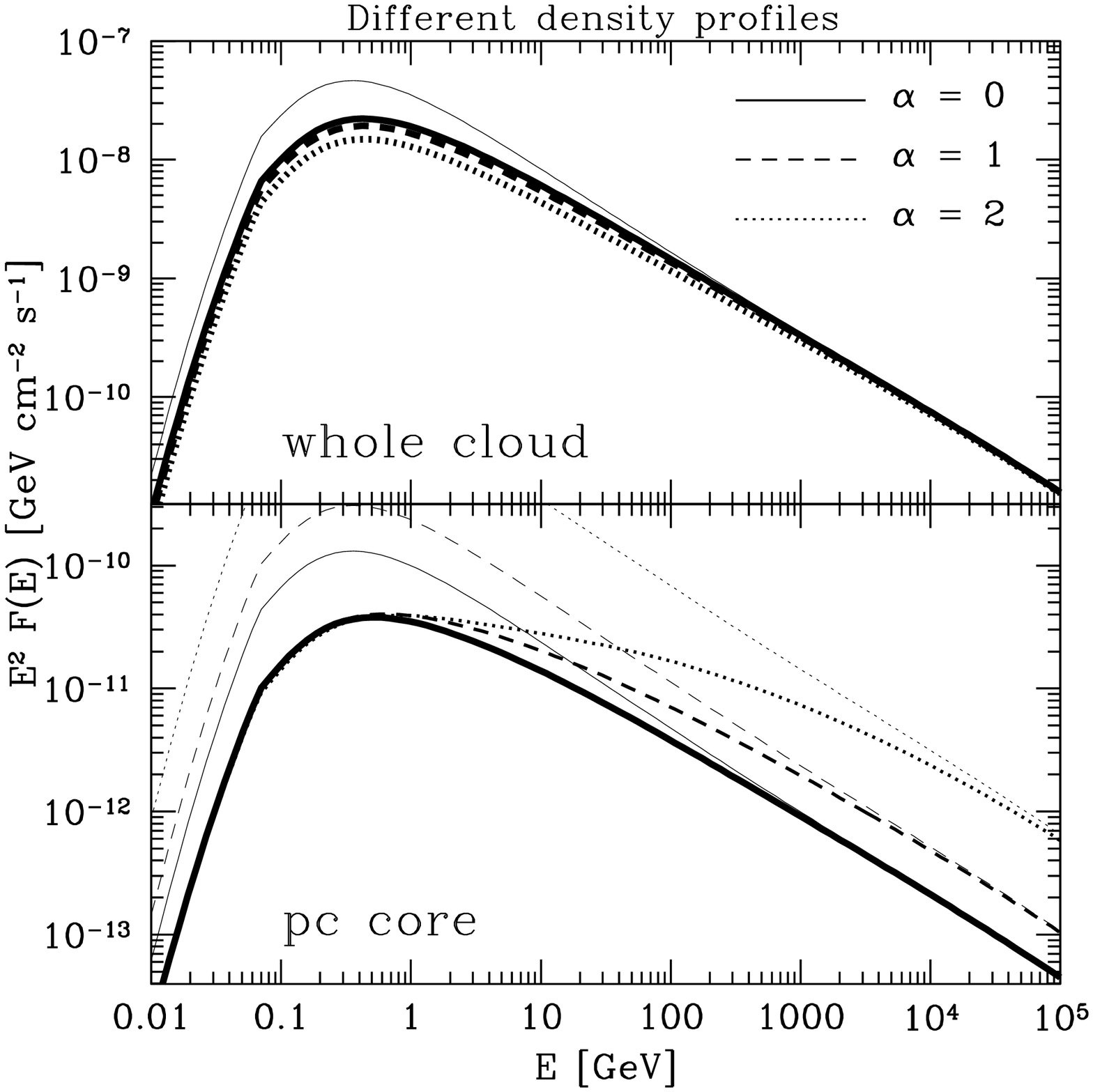}%{gammacoreBW.ps}
\caption{{\bf LEFT PANEL}: gamma ray emission from a cloud with $M = 2 \times 10^5 M_{\odot}$, $R_{cl} = 20$ pc and a flat density profile. The cloud distance is 1 kpc. Thick lines: $\pi^0$--decay gamma rays, thin lines: Bremsstrahlung gamma rays. {\bf RIGHT PANEL}: Thick lines: gamma ray emission (only $\pi^0$--decay component) from clouds with different density profiles and $\chi = 0.01$. Thin lines show the gamma ray emission one would observe if cosmic rays could freely penetrate the cloud.}
\label{fig:2}       % Give a unique label
\end{figure*}

Fig. \ref{fig:2} (left panel) shows the gamma ray spectra for a cloud of mass $M = 2 \times 10^5 M_{\odot}$ and radius $R_{cl} = 20 pc$. A flat density profile is assumed for a cloud at a distance of 1 kpc. The thick lines represent the contribution to the total gamma ray emission from $\pi^0$--decay (calculated following \cite{kelner}), while the thin lines represents the Bremsstrahlung contribution. Calculations have been performed adopting a diffusion coefficient with $\delta = 0.5$, but assuming three different values for $\chi = 1, 0.01, 0.001$ (curves top to bottom respectively). The Bremsstrahlung contributions become significant only below photon energies of $\sim 100 MeV$. 

The solid line represents the situation in which cosmic rays can freely penetrate the cloud, namely $\chi = 1$ (no suppression of the diffusion coefficient). This gives the maximum possible gamma ray luminosity for a passive cloud immersed in the galactic cosmic ray sea. If the diffusion coefficient is suppressed (dashed and dotted lines, corresponding to $\chi = 0.01, 0.001$) the total gamma ray luminosity is reduced, especially at low energies ($\sim 1$ GeV), while at high energies the canonical spectrum for a passive cloud filled by cosmic rays is recovered. This reflects the fact that high energy cosmic rays can freely penetrate inside clouds.
At $\sim 1$ GeV, the suppression of the gamma ray flux is roughly a factor of $\sim 2$ for $\chi = 0.01$ and a factor of $\sim 5$ for $\chi = 0.001$. Moreover, also the shape of the spectrum in modified, appearing flatter for lower values of $\chi$ (stronger suppression of diffusion). This fact may have very important implications for future GLAST observations of molecular clouds. In particular, the observation of gamma ray spectra harder than the ones expected from an isolated (passive) cloud pervaded by galactic cosmic rays can be interpreted in two different ways: \textit{i)} the galactic cosmic ray spectrum at the location of the cloud is different than the one measured locally; \textit{ii)} propagation effects inhibit the penetration of low energy cosmic rays in the cloud, making the resulting gamma ray spectrum harder. 

%can not be straightforwardly interpreted as an evidence of the non-universality of the galactic cosmic ray spectrum, since also the propagation effects described here might flatten the spectrum.

Molecular clouds constitute also potential targets for Cherenkov telescopes arrays, operating at photon energies greater than $100$ GeV.
The apparent angular size of a cloud with radius $R_{cl} \sim 20$ pc located at $d = 1$ kpc is $\vartheta_{cl} \sim 2 R_{cl}/d \sim 2.2^o$. This is significantly smaller than the telescope field of view (e. g. $\sim 5^o$ for HESS) and much larger than its angular resolution ($\sim 0.1^o$), thus Cherenkov telescopes can effectively map the gamma ray emission from clouds. Notably, a core with radius of half parsec will subtend an angle comparable with the telescope angular resolution. 
For this choice of the parameters, the gamma ray flux (only $\pi^0$--decay contribution) for a cloud of mass $2 \times 10^5 M_{\odot}$ is shown in Fig. \ref{fig:2} (right panel). The diffusion coefficient is the galactic one suppressed by a factor of 100 and the parameter $\alpha$ describing the slope of the density profile is varied. In the top panel the emission from the whole cloud is shown. Thick lines represent the spectrum for different values of $\alpha$: 0 (solid), 1 (dashed) and 2 (dotted). The thin line represents the spectrum one would observe if cosmic rays could freely penetrate the cloud. It can be seen that the emission from the whole cloud is not depending strongly on the density profile. The situation is much different if one considers the radiation from the inner parsec region (Fig. \ref{fig:2}, bottom panel), which in this particular case correspond with the radiation received within one angular resolution of the Cherenkov telescope. In this case the spectrum is strongly dependent on the assumption made on the density profile, especially at very high energies above $\sim 10$ GeV. This is because, for peaked density profiles ($\alpha > 0$) the exclusion of cosmic rays from cloud cores is much more effective with respect to the case of a flat profile ($\alpha = 0$), due to the enhanced energy losses and reduced diffusion there. This makes the gamma ray spectrum harder. On the other hand, since the cores are very dense, the suppression of the cosmic ray density is compensated by the higher efficiency of production of gamma rays. This explains why a higher level of the gamma ray emission from the densest cores is predicted.

Thus, we arrive at the important conclusion that \textit{the shape of the gamma ray spectrum from molecular cloud cores is determined not only by the diffusion properties of cosmic rays but also by the actual shape of the cloud density profile}.

\begin{figure*}
\centering
  \includegraphics[width=0.45\textwidth]{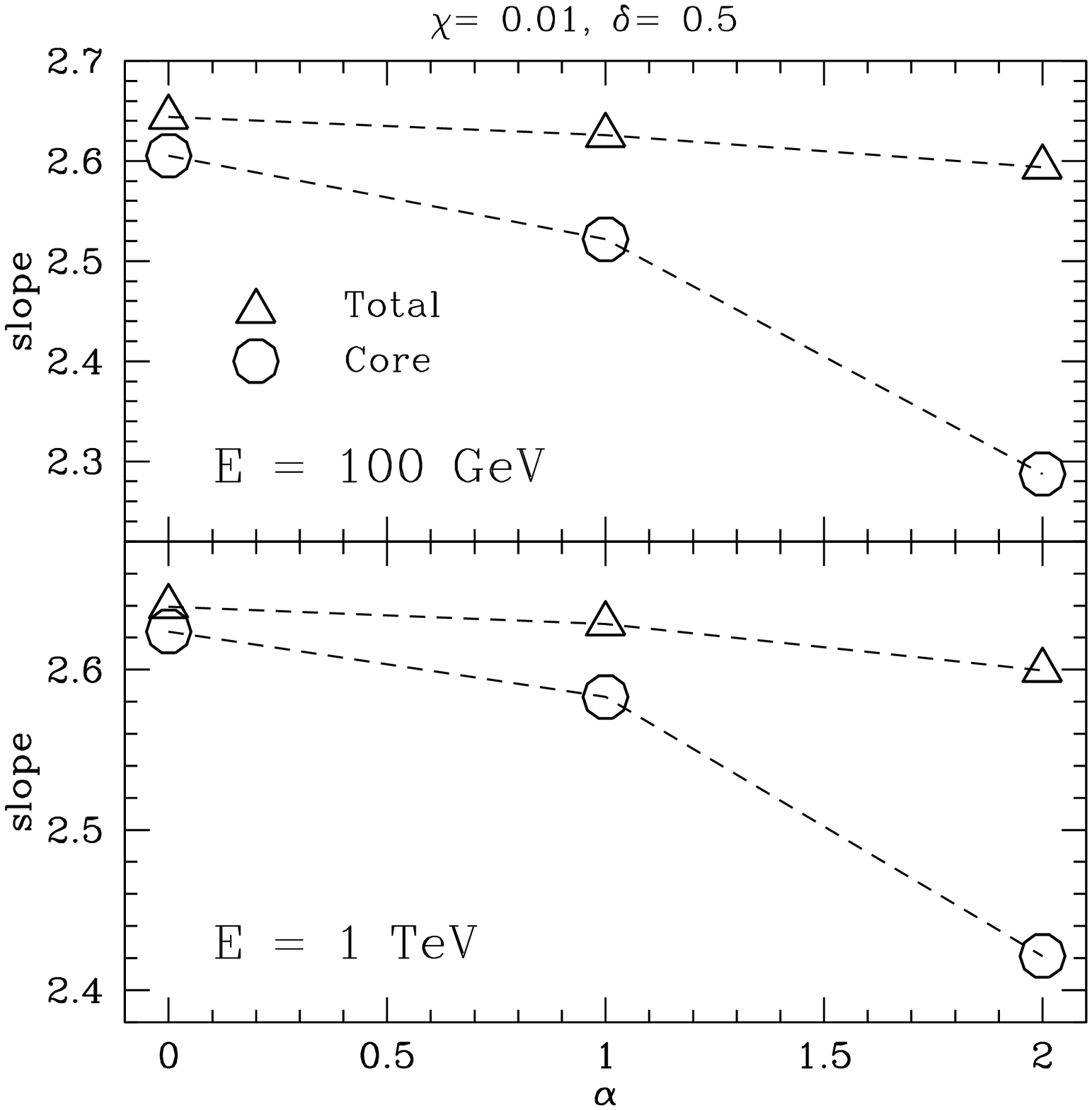}%{slopesBW.ps}
  \hspace{0.5cm} 
  \includegraphics[width=0.45\textwidth]{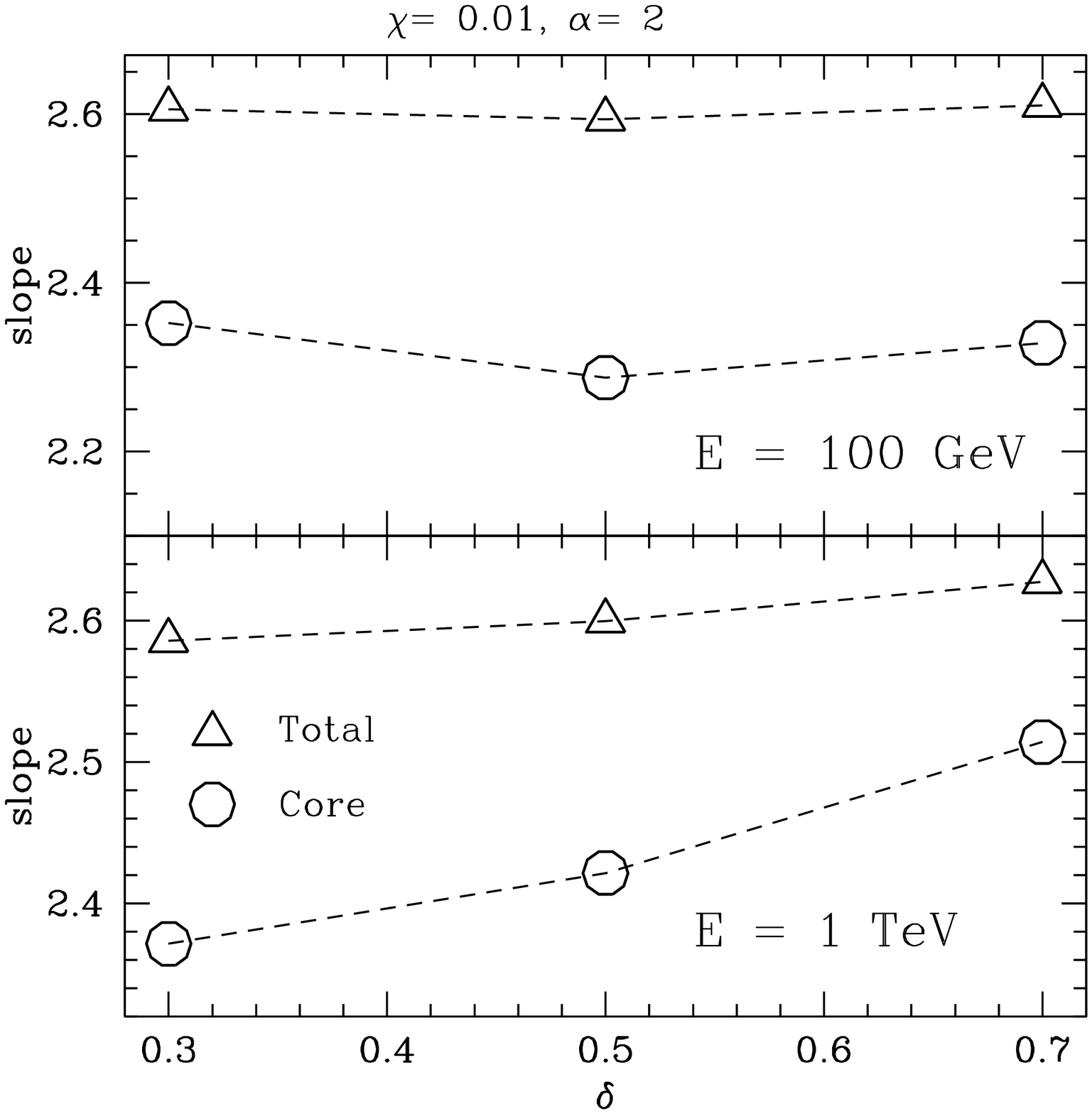}%{slopesBW2.ps}
\caption{Slope of the gamma ray spectrum for different sets of parameters. Triangles refer to the spectrum of the whole cloud, circles refer to the spectrum observed by looking at the cloud core.}
\label{fig:3}       % Give a unique label
\end{figure*}

This statement is confirmed in Fig. \ref{fig:3}, where the slope of the gamma ray spectrum at photon energies of $100$ GeV (top panels) and $1$ TeV (bottom panels) is plotted for different combinations of model parameters.
In the left panel we fix $\delta = 0.5$ but we change the slope of the density profile $\alpha$, while in the right panel, $\alpha$ is kept fixed and the energy dependence of the diffusion coefficioen $\delta$ is varied. In both plots $\chi = 0.01$. Triangles refer to the spectrum from the whole cloud and circles refer to the core. Again, we emphasize that, although the total gamma ray spectrum from the whole cloud is not strongly affected by the choice of the model parameters, the cloud cores may show a great variety of energy spectra.
As a consequence, gamma ray observations in the $TeV$ energy domain can be very useful to constrain parameters. In principle, if the cloud density profile could be extracted from independent observations (e.g. CO line emission) the shape of the gamma ray spectrum would tell us important information about the diffusion coefficient. Of course, the determination of the density profile is a extremely difficult task, since clouds show very complex structures. %However, the fact that the approach proposed above might work at least in idealized situations seems encouraging. 

\section{Conclusions}

In this paper we considered a giant molecular cloud embedded in the diffuse galactic cosmic ray flux. Assuming that the cosmic ray propagation inside the cloud proceeds in the diffusive regime, we studied the exclusion/penetration of cosmic rays into molecular clouds. Results can be summarized as follows:

\begin{itemize}

\item[$\bullet$]{if the diffusion coefficient inside the cloud is equal to the measured galactic one, cosmic rays can freely penetrate the cloud and the resulting high energy gamma ray emission has a spectral shape that closely resembles the one of the galactic cosmic rays.}

\item[$\bullet$]{the gamma ray emission above 100 MeV is dominated bi $\pi^0$--decay. Bremsstrahlung emission from secondary electrons is relevant only at smaller energies.}

\item[$\bullet$]{if the diffusion coefficient is suppressed with respect to the galactic one, cosmic rays can be effectively excluded from clouds. In particular, for a suppression of the diffusion coefficient of a factor of $\sim 100$, the exclusion becomes relevant at energies of tens-hundreds of GeV.}

\item[$\bullet$]{the exclusion of cosmic rays from clouds results in a suppression of the gamma ray flux, especially at $\sim$ GeV energies. This can have important consequences for the forthcoming GLAST observations.}

\item[$\bullet$]{Cherenkov telescopes such as HESS or VERITAS has the capability to map the gamma ray emission from clouds and to resolve the pc-scale cores. In particular, the shape of the gamma ray spectra from cloud cores may strongly depend on both the diffusion properties of cosmic rays and the shape of the cloud density profile. The effect of the density profile in shaping the gamma ray spectrum of molecular clouds was never considered before.}

\end{itemize}

This work has been stimulated by the recent detection of some molecular clouds in the galactic center region by HESS \cite{HESS} and by the forthcoming observations of the galactic plane by GLAST. It is beyond any doubt that future observations of molecular clouds by these two instruments, covering a broad energy interval from GeV to multi TeV photon energies, will provide deep insight into the problem of the origin of galactic cosmic rays.

\begin{acknowledgements}
SG acknowledges support from the Humboldt foundation.
\end{acknowledgements}

% BibTeX users please use
%\bibliographystyle{spmpsci}


\end{document}